\documentclass[%
 aip,
 rsi,
 superscriptaddress,
 amsmath,amssymb,
 reprint,%
]{revtex4-2}

\usepackage{graphicx}
\usepackage{dcolumn}
\usepackage{bm}

\usepackage[utf8]{inputenc}
\usepackage[T1]{fontenc}
\usepackage{mathptmx}
\usepackage{etoolbox}

\makeatletter
\def\@email#1#2{%
 \endgroup
 \patchcmd{\titleblock@produce}
  {\frontmatter@RRAPformat}
  {\frontmatter@RRAPformat{\produce@RRAP{*#1\href{mailto:#2}{#2}}}\frontmatter@RRAPformat}
  {}{}
}%

\usepackage{xcolor}
\usepackage{hyperref}
\hypersetup{colorlinks=true, allcolors=cyan}

\newcommand{\cyan}[1]{\textcolor{cyan}{#1}}
\newcommand{\eps}{$\varepsilon_{\mathrm{r}}^{\mathrm{He}}(T,P)$}

\usepackage[english]{babel}
\usepackage{wasysym}
\usepackage{amssymb}
\usepackage{changes}

\makeatother
\begin{document}

\preprint{AIP/123-QED}

\title[Advanced technique for measuring relative length changes under control of temperature and helium-gas pressure]{Advanced technique for measuring relative length changes under control of temperature and helium-gas pressure}

\author{Y. Agarmani} 
	\email[The author to whom correspondence may be addressed: ]{agarmani@physik.uni-frankfurt.de}
	
\affiliation{Institute of Physics, Goethe University Frankfurt, Max-von-Laue-Straße 1,\\ \,\,\,\,60438 Frankfurt am Main, Germany}

\author{S. Hartmann}%
\author{J. Zimmermann}%
\affiliation{Institute of Physics, Goethe University Frankfurt, Max-von-Laue-Straße 1,\\ \,\,\,\,60438 Frankfurt am Main, Germany}
\author{E. Gati}
\affiliation{Institute of Physics, Goethe University Frankfurt, Max-von-Laue-Straße 1,\\ \,\,\,\,60438 Frankfurt am Main, Germany}
\affiliation{Max-Planck-Institute for Chemical Physics of Solids, Nöthnitzer Straße 40,\\ \,\,\,\,01187 Dresden, Germany}
\author{C. Delleske}%
\author{U. Tutsch}%
\author{B. Wolf}%
\author{M. Lang}%
\affiliation{Institute of Physics, Goethe University Frankfurt, Max-von-Laue-Straße 1,\\ \,\,\,\,60438 Frankfurt am Main, Germany} 


\date{\today}


\begin{abstract}
We report the realization of an advanced technique for measuring relative length changes $\Delta L/L$ of mm-sized samples under control of temperature ($T$) and helium-gas pressure ($P$). The system, which is an extension of the apparatus described in Manna \textit{et al}., Rev. Sci. Instrum. \textbf{83}, 085111 (2012), consists of two $^4$He-bath cryostats each of which houses a pressure cell and a capacitive dilatometer. The interconnection of the pressure cells, the temperature of which can be controlled individually, opens up various modes of operation to perform measurements of $\Delta L/L$ under variation of temperature and pressure. Special features of this apparatus include the possibilities (1) to increase the pressure to values much in excess of the external pressure reservoir, (2) to substantially improve the pressure stability during temperature sweeps, (3) to enable continuous pressure sweeps both with decreasing and increasing pressure, and (4) to simultaneously measure the dielectric constant of the pressure-transmitting medium helium, $\varepsilon_{\mathrm{r}}^{\mathrm{He}}(T,P)$, along the same $T$-$P$ trajectory as used for taking the $\Delta L(T,P)/L$ data. The performance of the setup is demonstrated by measurements of relative length changes $(\Delta L/L)_T$ at $T=180$\,K of single crystalline NaCl upon continuously varying the pressure in the range $6\,\mathrm{MPa}\leq P \leq 40$\,MPa.   
\end{abstract}


\maketitle

\section{Introduction} \label{sec:Introduction}

The coefficient of thermal expansion, describing how the size of a material changes upon changing the temperature, is a fundamental thermodynamic quantity.\,\,In fact, measurements of relative length changes, $\Delta L_i(T)/L_i = [ L_i(T)-L_i(T_0) ]/L_i(T_0)$, with $L_i$ the sample length along the axis $i$ and $T_0$ a reference temperature, have proven a powerful tool for exploring the electronic, magnetic and lattice properties and their directional dependencies. These investigations have been of great importance for studying cooperative phenomena such as phase transformations or crossovers and their variations with external parameters such as temperature and magnetic fields. Among the various methods for measuring length changes of a sample, capacitive dilatometry stands out due to the extraordinarily high resolution of $\Delta L/L \geq 10^{-10}$ which exceeds other techniques, including X-ray diffraction\cite{Barron1980}, optical interferometers \cite{Hamann2019} and strain gauges \cite{Kabeya2011}, by at least one order of magnitude. A first implementation for dilatometry under high pressure using a capacitive technique was presented in Ref. \onlinecite{Fietz2000}. In a recent work, we reported on the development of an apparatus, which enables to expand the range of application of capacitive dilatometry to finite hydrostatic pressure \cite{Manna2012}, providing a more sensitive alternative to commonly used methods for dilatometric studies under pressure such as X-ray diffraction and strain gauges \cite{Sakai1999, Gati2021}. The latter two techniques along with ultrasonic studies \cite{Anderson1966} are also used to measure the compressibility of solids (see Ref.\,\onlinecite{Decker1972} for a review), albeit with reduced sensitivity. The setup described in Ref.\,\onlinecite{Manna2012} is based on the use of a high-resolution capacitive dilatometer in combination with a variable helium-gas pressure environment for measuring relative length changes under control of temperature and hydrostatic pressure. This technique was successfully applied in measurements of $\Delta L(T,P)/L$ on an organic charge-transfer salt, which is located close to the Mott metal-insulator transition, for investigating the coupling of the critical electronic system to the lattice degrees of freedom \cite{Gati2016}. The results obtained by Gati \textit{et al.}, disclosing a pronounced non-linear strain-stress relation on approaching the critical endpoint of the Mott transition \cite{Gati2016}, has demonstrated the high potential of dilatometric studies under variable helium-gas pressure. This technique is particularly useful for those investigations where a fine and \textit{in situ} pressure tuning is demanded. In general, two modes of operation are of interest including (A) temperature sweeps over a sufficiently wide range of temperatures at $P \approx$ const. conditions, as well as (B) pressure sweeps at $T\approx$ const. conditions. Depending on the problem under investigation, these experiments may pose challenges on the process control. 
\begin{figure*} [t]
	\centering
	\includegraphics[width=0.8\textwidth]{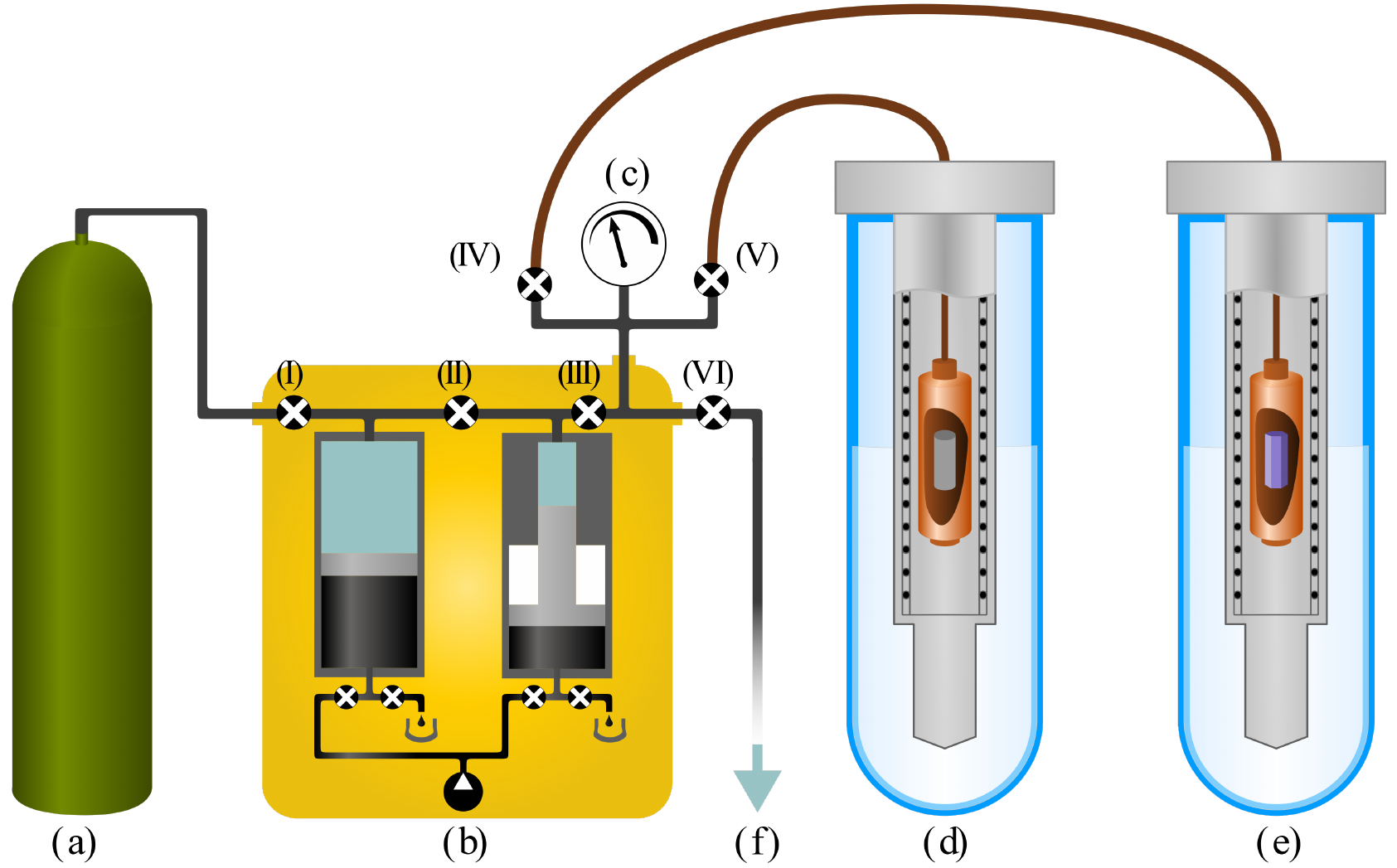}
	\caption{\label{fig:complet_setup}Sketch of the setup with all relevant components. A gas bottle (a), loaded with helium gas at $P \leq30$\,MPa, serves as a pressure reservoir. The gas is injected into a two-stage compressor unit (b). A digital manometer (c) is used to determine the pressure. The compressor is connected via CuBe capillaries to the sample- (d) and reference-setup (e) each of which consists of a $^{4}$He-bath cryostat and an insert including a pressure cell and a dilatometer cell. The pressure loading of the pressure cells in (d) and (e) can be controlled individually through the valves (IV) and (V). The pressure can be released into the recovery line (f) via valve (VI).}
\end{figure*}
This includes the stabilization of pressure while ramping the temperature up and down (A) as well as a sufficiently good temperature control in studies where the pressure is varied (B). The latter also comprises the need to change the pressure in a continuous way at a desired and sufficiently low rate to ensure thermal equilibrium. Moreover, in specific cases, measurements at both decreasing and increasing pressure at $T\approx$\,const. could be of interest, in particular for studying pressure-induced phase changes. The setup described here, consisting of two interconnected pressure cells each of which is equipped with its own temperature-control unit, considerably improves the range of applications and the quality of the pressure- and temperature control as compared to a system with a single measuring cell and a gas reservoir controlled at room temperature (cf. Ref. [5]). Here, the following challenges have been addressed successfully: 1) The maximum accessible pressure, $P_\mathrm{max}$, has been increased significantly by drastically reducing the gas volume at room temperature. In particular, $P_\mathrm{max}$ is no longer limited by the maximum pressure of the external pressure reservoir – a value which, due to safety regulations, strictly limits gas-containing parts at room temperature by placing a limit on the product of volume $\times$ pressure. 2) The unwanted change of pressure during a temperature sweep – an inevitable consequence of working with a closed system – has been drastically reduced. 3) Continuous temperature-driven pressure sweeps can be performed both with decreasing and increasing pressure for a wide range of sweep rates with a quality unmatched by any mechanical pressure control. 4) As the capacitive dilatometer is surrounded by the pressure-transmitting medium helium, a precise knowledge of its dielectric constant $\varepsilon_{\mathrm{r}}^{\mathrm{He}}(T,P)$ is required in the data analysis, since $\varepsilon_{\mathrm{r}}^{\mathrm{He}}$ varies significantly with temperature and pressure at low temperatures. Our setup allows for a simultaneous measurement of $\varepsilon_{\mathrm{r}}^{\mathrm{He}}(T,P)$ along the same $T$-$P$ trajectory as used for taking the $\Delta L(T,P)/L$ data.


\section{Interconnection of two pressure cells} \label{sec:INTERCONNECTION}

Figure \ref{fig:complet_setup} shows a sketch of the setup with all relevant components. The system consists of two He-bath cryostats (d) and (e) both of which are equipped with a pressure cell and a capacitive dilatometer. The cryostat (d) and its insert, referred to as sample setup from here on, are described in detail in Ref.\,\onlinecite{Manna2012}. This setup has been complemented by adding a second, almost identical system, referred to as a reference setup. Both pressure cells are connected via capillaries to a compressor unit (b) and a helium reservoir (a). The compressor unit is equipped with high-pressure valves (I) – (VI) for controlling the gas flow.

The helium gas, used as pressure-transmitting medium, is provided by utilizing a standard gas bottle (a) as a pressure reservoir. The gas bottle can be loaded with a maximum allowed pressure of 30\,MPa. The purity of the helium used is typically 99.999\,\%. The bottle is connected via valve (I) to the two-stage helium-gas compressor U11 300\,MPa (b), which is commercially available (for more details see Ref.\,\onlinecite{Manna2012}) and was developed in cooperation with the Institute of High Pressure Physics, Polish Academy of Sciences, Unipress Equipment Division, in the following abbreviated as Unipress. In order to reach pressures $P > 30$\,MPa, valve (I) needs to be closed and the two stages of the compressor pressurize the helium gas to the desired value. Whereas the first (low-pressure) stage enables pressures $P<70$\,MPa, the second (high-pressure) stage reaches pressures of $70$\,MPa $\leq P \leq 250$\,MPa. The high-pressure stage is connected via valve (III) to two CuBe capillaries of similar length (inner diameter 0.3\,mm, outer diameter 3\,mm), one of which is connected to the sample setup, the other one is connected to the reference setup. The pressure cells and capillaries are manufactured by Unipress. Valves (IV) and (V) allow for an individual pressure loading of the pressure cells. Valve (VI) enables to relief the pressure into the helium recovery system. 

The two identical pressure cells, used in the sample- and reference setups, are made of CuBe (see Refs.\,\onlinecite{Manna2012, Gati2016} for details) with an inner/outer diameter of 36\,mm/56\,mm and a maximum permissible pressure of 250\,MPa. Each pressure cell houses a capacitive dilatometer cell \cite{Pott1983} with the same design, manufactured by Kuechler Innovative Measurement Technology \cite{Kuechler2012}. The pressure cell of the sample setup (d) is opened regularly for mounting the sample to be investigated. On the other hand, the reference-pressure cell (e) is kept closed and is opened only for maintenance. A high-purity (99.99\,\%) aluminum sample of 4.5\,mm length is installed in the reference-dilatometer cell and serves as a reference material. The criteria used for selecting this material are (i) a known and well-reproducible thermal expansion and (ii) a small compressibility ensuring a negligibly small effect of pressure in the pressure range used here. A digital manometer LEO5 (KELLER AG) (c) is used to determine the pressure with a resolution of $\vert \Delta P \vert \leq 0.1\,$MPa for pressures up to 100\,MPa and $\vert \Delta P \vert \leq0.03$\,MPa for pressures up to 30\ MPa. The home-made inserts, used for the $^4$He cryostats (d) and (e), were designed with special focus placed on ensuring a good temperature control of the pressure cells. The essential construction elements, described in Ref.\,\onlinecite{Manna2012}, include the pressure cell, the capillary, the capillary heaters, the thermometers (Cernox®, LakeShore Cryotronics), and the inner vacuum can (IVC) made of copper (diameter 66\,mm). This can is wrapped with heating foil (Kapton®, thermofoil, Minco company) to ensure a homogeneous heat input into the pressure cells. The IVC is surrounded by an outer vacuum can made of stainless steel (diameter 75\,mm). Both cans are connected to pump lines that can be vented independently with $^4$He exchange gas. The IVCs of both inserts are filled with $^4$He gas at low pressure of typically $0\leq P \leq 0.01$\,MPa (at 300\,K) for ensuring a good thermal contact between the heater and the pressure cell, see Ref.\,\onlinecite{Manna2012} for details of the inserts including thermometry and temperature control unit.

The accessible temperature range for the inserts is $1.4$\,K$\leq T\leq300$\,K. The maximum accessible pressure $P_\mathrm{max}$ is constrained by the operating principle of the capacitive dilatometer, requiring a freely movable upper capacitor plate. Thus, $P_\mathrm{max}$ is limited by the solidification line $T_\mathrm{sol}^{\mathrm{He}}(T,P)$ of the pressure-transmitting medium helium, see Ref.\,\onlinecite{Mills1980} and references cited therein, which is 60\,MPa at 10\,K and 1.45\,GPa at 77\,K, for example.


\section{Accessible modes of operation} \label{sec:Modes of operation}

Depending on the physical problem under investigation and the desired temperature- and pressure range of the experiment, different modes of operation are accessible. For experiments under finite pressure, where the variations in temperature and pressure imply significant changes of the dielectric constant of helium, simultaneous measurements of \eps\ can be performed. This procedure will be described in section \ref{sec:OperationA}. If temperature sweeps at $P\approx$ const. conditions are demanded, special measures can be taken to keep temperature-induced pressure changes small, see section \ref{sec:OperationB}. In section \ref{sec:OperationC} we describe a mode of operation, which allows for pressure sweeps with both decreasing and increasing pressure. In the description of the various modes of operation, we use a simplified sketch of the relevant parts of the inserts of the sample- and reference setups as displayed on the right side of figure \ref{fig:complet_setup}. While the sample setup (Fig.\,\ref{fig:complet_setup}\,(d)) contains the dilatometer cell with the sample of interest, the reference setup (Fig.\,\ref{fig:complet_setup}\,(e)) contains the dilatometer cell with the aluminum sample installed. The valves (III) to (VI) allow for an individual control of the gas flow. Both setups are supplied with their own temperature-control units enabling to run independent temperature-time ($T$-$t$) profiles.


\subsection{Simultaneous determination of relative length change and the dielectric constant of the pressure-transmitting medium } \label{sec:OperationA}
\begin{figure*} [t]
	\centering
	\includegraphics[width=0.69\textwidth]{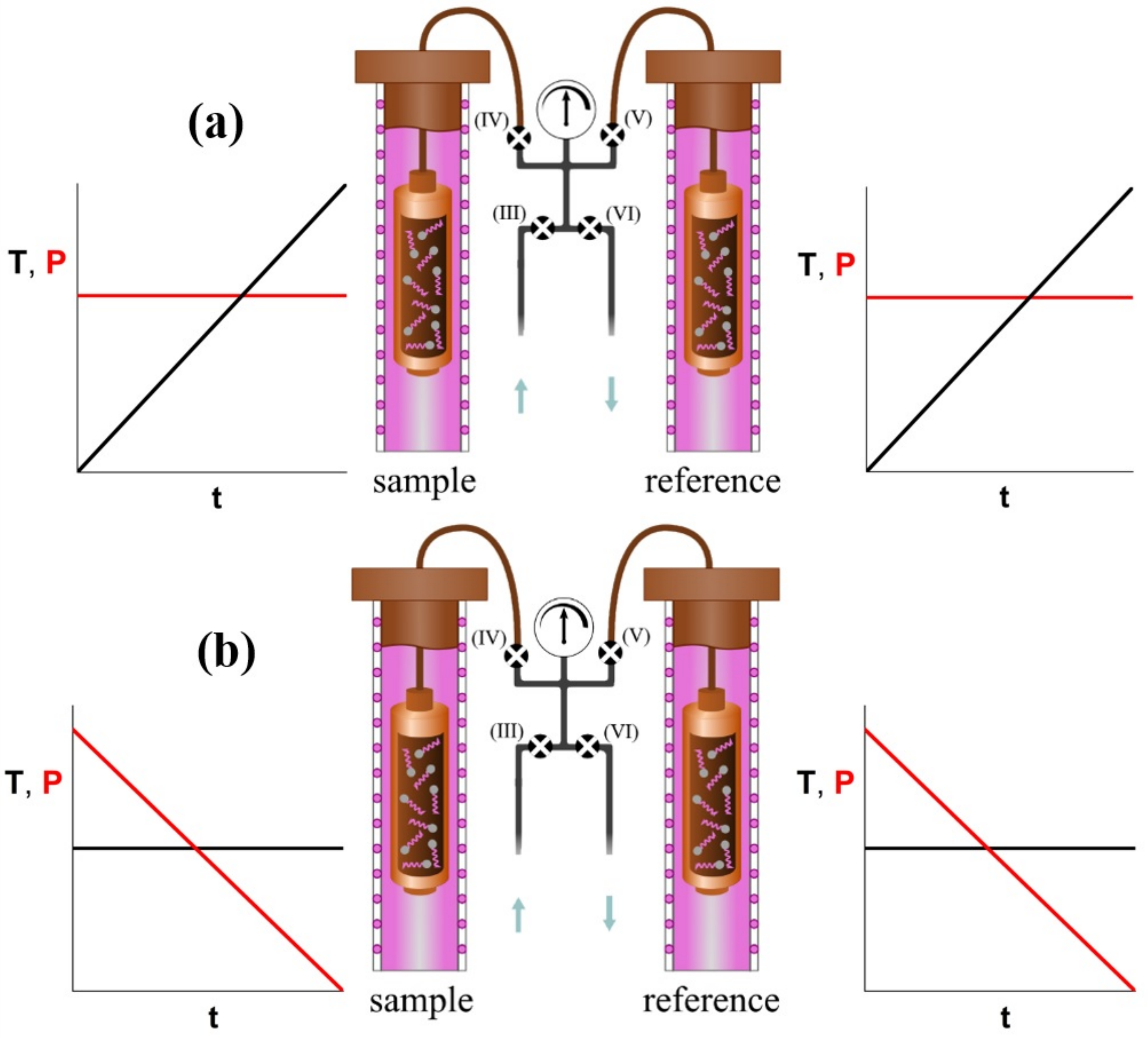}
	\caption{\label{fig:operation_A}Sketch of the mode of operation for a simultaneous determination of length changes of a sample $\Delta L(T,P)/L$ and the dielectric constant of the pressure-transmitting medium helium \eps. The temperature control units for both systems are operated in a way so that the temperature is the same for both setups, illustrated by the same color. The valves (IV) and (V) are open so that the pressure is the same in both pressure cells. For $T$-sweeps (a), the temperature in both setups is changed with the same rate $q_1 = q_2$, while $P$ is approximately held constant. For $P$-sweeps (b), the pressure of both setups is changed with the same rate, while $T$ is held constant, ensuring the same conditions for the sample- and reference-measurement.}
\end{figure*}
The pressure-transmitting medium helium in the pressure cell acts as dielectric between the capacitor plates of the dilatometer cell. It therefore affects the capacitance $C$ of the dilatometer and thus has to be taken into account in converting changes of the capacitance into relative length changes, see Eq.\,\ref{eq:capacitance_pressure} below and Ref.\,\onlinecite{Manna2012}. As there is only limited data on \eps\ available in the literature (see Refs.\,\onlinecite{Schmidt2003, Manna2012}), a simultaneous determination of both the relative length change of the sample $\Delta L(T,P)/L$ and \eps\ along the same $T$-$P$ trajectory brings major benefits. As will be shown below, \eps\ can be derived from measuring capacitance changes in the reference setup. In vacuum, the capacitance $C$ in the reference setup is described by:
\begin{equation} \label{eq:capacitance}
C(T,P=0)= \varepsilon_0 \frac{A}{d(T,P=0)}\ ,
\end{equation}
with the vacuum permittivity $\varepsilon_0$, the area of the capacitor plates $A$, and the distance between the plates $d$. If the system is under helium pressure $P$, this expression changes into:
\begin{equation} \label{eq:capacitance_pressure}
C(T,P)= \varepsilon_0 \varepsilon_{\mathrm{r}}^{\mathrm{He}}(T,P) \frac{A}{d(T,P)}\ ,
\end{equation}
with the dielectric constant of helium \eps. It is reasonable to assume that in the pressure range discussed here, i.e., $P\leq250$\,MPa, the area $A$ does not change significantly with pressure, see below for explanation. By combining Eq.\,\ref{eq:capacitance} and \ref{eq:capacitance_pressure} we obtain:
\begin{equation} \label{eq:dielectric_constant_1}
\varepsilon_{\mathrm{r}}^{\mathrm{He}}(T,P) = \frac{C(T,P)}{C(T,P=0)} \cdot \frac{d(T,P)}{d(T,P=0)}\ .
\end{equation}
The compressibility of the dilatometer cell (made of a copper beryllium alloy with a small beryllium concentration of 1.84\,\%\cite{Kuechler2012}) and the reference sample (aluminium sample installed in the reference setup) is rather small, with $\kappa_{\mathrm{CuBe}}(300\, \mathrm{K}) = 7.9 \cdot 10^{-6}$\,MPa$^{-1}$\,\cite{Kamarad2004} and $\kappa_{\mathrm{Al}}(300\,\mathrm{K}) = 13.85 \cdot 10^{-6}$\,MPa$^{-1}$\,\cite{Kittel2004}, respectively. Thus, the corresponding pressure-induced changes in both $d$ (caused by length changes of aluminum) and the dimensions of the dilatometer cell give rise to changes in the capacitance of $\mathcal{O}$($10^{-3}$\,pF) for capacitance of typically $C\approx13\,$pF. This has to be compared with changes in \eps\ throughout the ranges of temperature (1.4\,K$\leq T\leq 300$\,K) and pressure ($P\leq250$\,MPa) covered in the experiment, which cause changes in the capacitance of $\mathcal{O}$($10^{-1}$\,pF). It is therefore reasonable to set $\mathrm{d}(T,P=0)=\mathrm{d}(T,P)$ in Eq.\,\ref{eq:dielectric_constant_1}, yielding:
\begin{equation} \label{eq:dielectric_constant_2}
\varepsilon_{\mathrm{r}}^{\mathrm{He}}(T,P) = \frac{C(T,P)}{C(T,P=0)}\ .
\end{equation}
Thus, results obtained in measurements of $C(T,P)$ in the reference setup, combined with $C(T,P=0)$ data collected in an independent run at $P = 0$, allow to determine \eps\ simultaneously with the measurements of $\Delta L (T,P)/L$. 

A sketch of the corresponding idealized modes of operation is shown in figure \ref{fig:operation_A}. The valves (IV) and (V) are open so that the pressure equilibrates in both pressure cells. For $T$-sweeps (a), both pressure cells are stabilized at the same starting temperature $T_{\mathrm{start}}$. Then the temperature is varied in the same way for both setups, normally by applying a linear $T$-$t$ profile as sketched in figure \ref{fig:operation_A}\cyan{(a)}. A typical ramp rate for such thermal expansion measurements is 1.5\,K/h. Ideally, the pressure should stay constant during a temperature sweep. In reality, however, the pressure varies by an amount $\Delta P$ depending on the $T$-$P$ range of operation, see section \ref{sec:OperationB}. For $P$-sweeps, shown in Fig.\,\ref{fig:operation_A}\,\cyan{(b)}, both pressure-cells are hold at the same constant temperature and the pressure is released [increased] by opening valve (VI) [(III)]. The change in pressure is the same for both pressure cells. No matter which option is chosen, the $C(T,P)$ data for the reference setup are collected simultaneously with the $\Delta L(T,P)/L$ data for the sample under investigation.


\subsection{Temperature sweeps with improved stabilization of pressure} \label{sec:OperationB}

\begin{figure*} [t]
	\centering
	\includegraphics[width=0.72\textwidth]{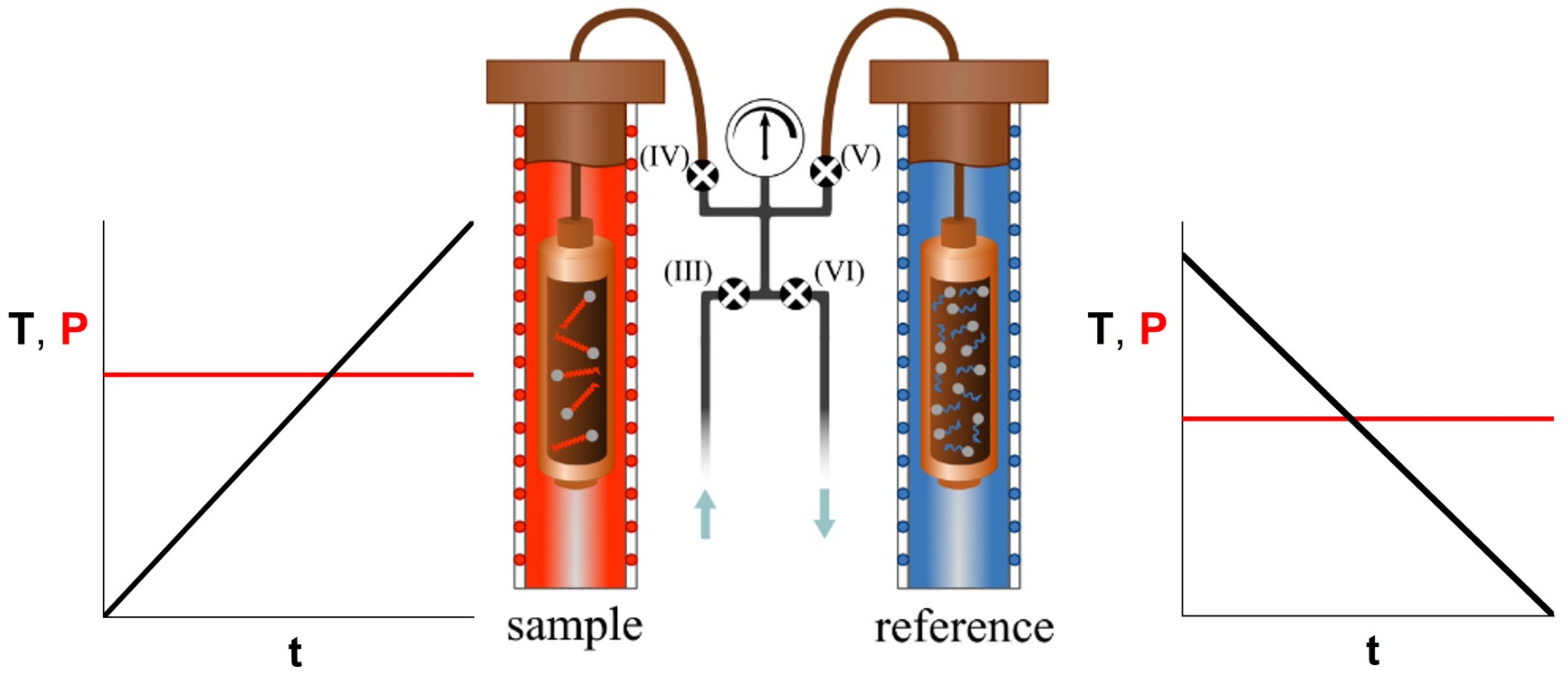}
	\caption{\label{fig:operation_B}Sketch of the mode of operation for improving the pressure stability during temperature sweeps. While the temperature is increased from $T_1$ to $T_2$ in the sample-pressure cell (illustrated by the red color in the IVC) at a rate $q_1$, the temperature in the reference-pressure cell is decreased from $T_2$ to $T_1$ with $q_2 = -q_1$ (illustrated by the blue color in the IVC). }
\end{figure*}
The mode of operation described in this section reduces unwanted changes of pressure during temperature sweeps. For a closed system, where the volume and the number of He-particles are kept constant, the ideal gas law implies that a change in the temperature $\Delta T$ gives rise to a change in pressure $\Delta P \propto \Delta T$. The volumes involved include the gas bottle (a) (cf.\,\,Fig.\,\ref{fig:complet_setup}) $V_{\mathrm{bottle}}=5 \cdot 10^4$\,cm$^3$, the first- and second stage of the compressor (b) with maximum volumes of $V_{1^{\mathrm{st}},\,\mathrm{max}}=720$\,cm$^3$ and $V_{2^{\mathrm{nd}},\,\mathrm{max}}=103$\,cm$^3$, respectively, and the pressure cells $V_{\mathrm{cell}}=81.4$\,cm$^3$, each of which houses a dilatometer cell $V_{\mathrm{dil}} \sim 20$\,cm$^3$. Both pressure cells are connected to the compressor by a capillary of length $l\sim 10$\,m with an inner diameter of 0.3\,mm. For measurements at $P\leq 30$\,MPa, the gas bottle remains connected to the system and serves as a large-volume room-temperature gas reservoir. The entire multi-component system can be considered to be in equilibrium ensuring the same pressure in all components of the system. As the volume of the gas bottle is much bigger than the volume of the pressure cell, $V_{\mathrm{bottle}} \approx 600 \cdot V_{\mathrm{cell}}$, the pressure changes within the cell, induced by changes of its temperature, are negligibly small.\, 

\begin{figure} [b]
	\centering
	\includegraphics[width=0.47\textwidth]{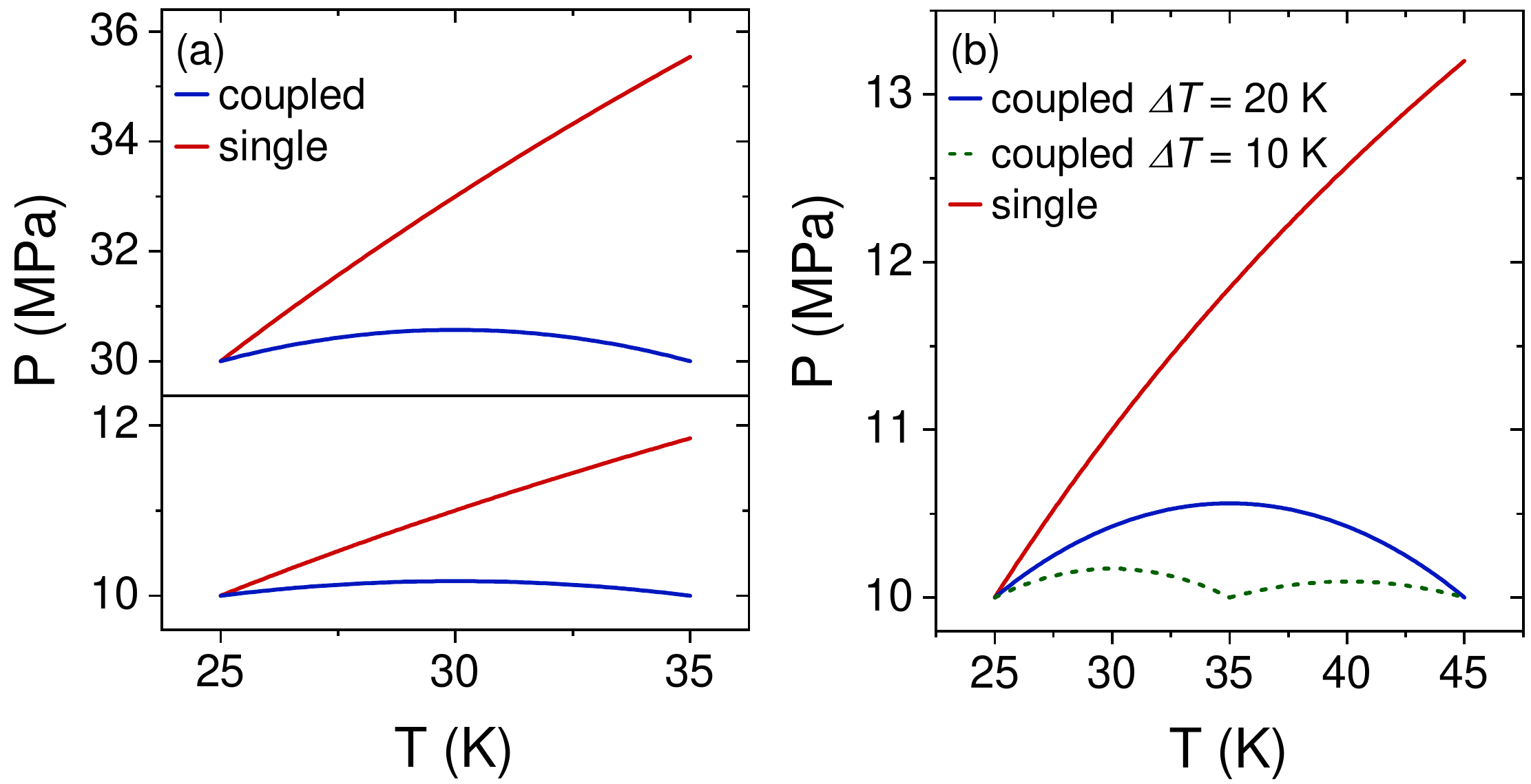}
	\caption{\label{fig:operation_B_model}Simulations of the variation of pressure during a temperature sweep. For details of the simulations see main text. The red lines represent the change in pressure when only one pressure cell (single) is connected to the compressor unit, corresponding to the standard setup. The blue lines correspond to pressure changes when both pressure cells (coupled) are coupled and connected to the compressor unit while running a process shown in figure \ref{fig:operation_B}. (a) Variation in pressure for two different target pressures: The upper panel presents the calculation for $P = 30$\,MPa while the lower panel shows the pressure changes for $P = 10$\,MPa. (b) Variation in pressure for a target pressure of 10 MPa while sweeping from $T_1 = 25$\,K to $T_2 = 45$\,K for different operation modes, i.e., traversing the temperature interval in a single step of width $\Delta T = 20$\,K, or in two steps of width $\Delta T = 10$\,K (in green), as indicated in the figure.}
\end{figure}

For measurements at $P > 30$\,MPa, the maximum allowed pressure for the gas bottle, valve (I) is closed and the two stages of the compressor are being activated. The higher the target pressure is, the smaller becomes the remaining volume of the pressurized system, i.e., $V_{1^{st}}\leq V_{1^{\mathrm{st}},\,\mathrm{max}}$ and $V_{2^{\mathrm{nd}}} \leq V_{2^{\mathrm{nd}},\,\mathrm{max}}$. Thus, with increasing pressure, the temperature-induced pressure changes become more and more significant. This effect is further reinforced with decreasing the temperature. When high-precision measurements of $\Delta L(T,P)/L$ over a limited range of temperature $T_1 \leq T \leq T_2$ are demanded, where only small variations in pressure can be tolerated, the reference setup can be used for stabilizing the pressure. The corresponding mode of operation is sketched in figure \ref{fig:operation_B}. Prior to the measurements, the sample setup is stabilized at a temperature $T_{\mathrm{start}} = T_1$ whereas the reference setup is stabilized at a temperature $T_{\mathrm{end}} = T_2$. The pressure cells are kept connected via the open valves (IV) and (V) to ensure an exchange of gas particles. On starting the measurement sequence, the temperature of the sample setup is increased in a linear $T$-$t$ fashion from $T_1$ to $T_2$ with the desired ramp rate $q_1=\Delta T/\Delta t$. This process is accompanied by simultaneously decreasing the temperature of the reference system from $T_2$ down to $T_1$ at a rate $q_2$ = $-q_1$. With this mode of operation, the temperature-induced changes in the sample pressure cell can be significantly reduced. Simulations of this effect based on the ideal gas law are shown in figure \ref{fig:operation_B_model}. In the calculations, the volume of the pressure cells and the maximum available volume of the two stages of the compressor are taken into account, whereas the volume of the capillaries has been neglected.
 
\begin{figure*} [!t]
	\centering
	\includegraphics[width=0.69\textwidth]{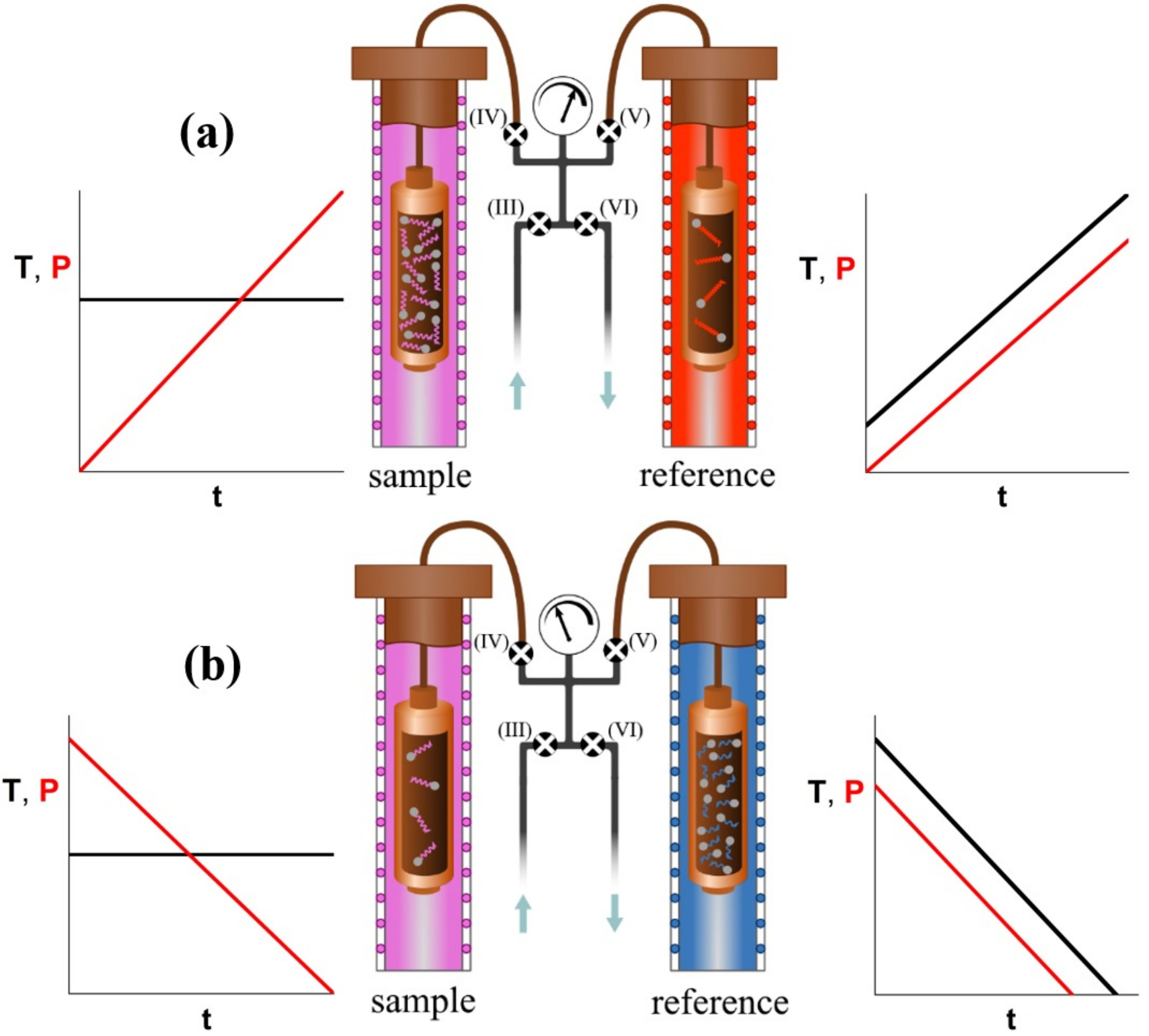}
	\caption{\label{fig:operation_C}(a) Sketch of the mode of operation for performing pressure sweeps (with increasing pressure) at $T =$ const. in the sample-pressure cell. To this end the reference-pressure cell is warmed up at a rate q$_2$ resulting in an overall increase of pressure in both pressure-cells. (b) Reversed process: The sample-pressure cell is held at constant $T$ while the reference-pressure cell is cooled down at a rate $-q_2$  resulting in an overall decrease of pressure in both pressure cells.}
\end{figure*}
Figure \ref{fig:operation_B_model}\,\cyan{(a)} shows the calculated change in pressure resulting from sweeping the temperature from 25\,K to 35\,K at two different starting pressures of 10\,MPa (lower panel) and 30\,MPa (upper panel). The red curves correspond to the standard operation mode where only one pressure cell (single) containing the sample is connected to the compressor giving rise to a significant increase of $\Delta P \sim 2$\,MPa (at an initial pressure of 10\,MPa) and $\Delta P \sim 6$\,MPa (at 30\,MPa). This can be compared to simulations of $\Delta P$ (blue curve) for the operation mode described above, where both cells (coupled) are connected to the compressor unit and their temperatures are swept at opposite rates $q_2 = -q_1$ as indicated in figure \ref{fig:operation_B}. Here, the corresponding $P$–$T$ curves (blue curves) are characterized by a dome shape with $P(T_1) = P(T_2)$ and a maximum pressure change of only $\Delta P \sim 0.2$\,MPa (at 10\,MPa) and $\Delta P \sim 0.6$\,MPa (at 30\,MPa). This mode of operation thus indicates a massive improvement in the stabilization of pressure during temperature sweeps. Figure \ref{fig:operation_B_model}\,\cyan{(b)} illustrates how the performance can be even further improved. The figure compares the variations in pressure, intended to be stabilized at 10\,MPa, for three different modes of operation while sweeping the temperature from $T_1 = 25$\,K to $T_2 = 45$\,K. 
The red line indicates the pressure increase by using a single pressure cell, yielding an increase of $\Delta P\sim 3$\,MPa. This effect can be reduced to $\Delta P \sim 0.6$\,MPa by the simultaneous use of two pressure cells as described above. The $\Delta P$ can be even further reduced by dividing the temperature interval ($T_1$, $T_2$) in half, and performing consecutive runs in both subsections (dotted green line). The corresponding maximum pressure change amounts to only $\Delta P \sim 0.2$\,MPa.


\subsection{Pressure sweeps at an adjustable sweep rate} \label{sec:OperationC}

In the previous version of the setup described in Ref.\,\onlinecite{Manna2012}, consisting of a single pressure cell connected to the compressor unit and gas bottle, continuous pressure sweeps were limited to a maximum pressure of $P_{\mathrm{max}} = 30$\,MPa, the maximum allowed pressure for the gas bottle. In addition, measurements could be performed only on decreasing the pressure (by slightly opening valve (VI)), which implied a nonlinear $P$–$t$ profile. As sketched in figure \ref{fig:operation_C}, the use of two interconnected setups opens up new possibilities for performing pressure sweeps at an adjustable sweep rate including an increased $P_{\mathrm{max}}$. In the initial stage of the process, the pressure cells are connected via the open valves (IV) and (V) and are loaded with an initial pressure $P_1$ from the compressor unit. The connection to the compressor and backline, valves (III) and (VI), is then closed. While the sample-pressure cell is held at a constant temperature, the reference-pressure cell is warmed up with a certain rate $q_2$. This process is accompanied by a smooth increase of pressure in the coupled system. Likewise, decreasing the temperature in the reference-pressure cell at a rate of $-q_2 $ induces a continuous decrease in pressure. The choice of the process parameters strongly depends on the targeted temperature and pressure range. In general, the lower the temperature and the higher the pressure of the sample are, the wider is the temperature change the reference-pressure cell has to pass. The possibility of performing pressure sweeps with increasing and decreasing pressure enables for example detailed investigations of phase transitions, in particular studying potential hysteresis effects. In addition, the interconnected two-pressure-cell system, which can be run independently from the compressor unit and gas bottle, enables controlled pressure sweeps up to pressures significantly higher than 30\,MPa.


\section{TEST MEASUREMENTS OF THE VARIOUS MODES OF OPERATION } \label{sec:Test_modes_of_Operations}

In the previous paragraph \ref{sec:Modes of operation} we have described various modes of operations offering unique possibilities for performing experiments for which special and well-defined conditions are demanded. As the scenarios described there consider idealized cases, the performance under real conditions may show some deviations the size of which depend on the chosen parameter range. In this paragraph we describe the results of test measurements for the different modes of operation. For selected cases we compare the performance with model calculations for the idealized system.


\subsection{Temperature sweeps at $P \approx$ const. conditions } \label{sec:Tsweep_at_P_const}

\begin{figure} [!h]
	\centering
	\includegraphics[width=0.48\textwidth]{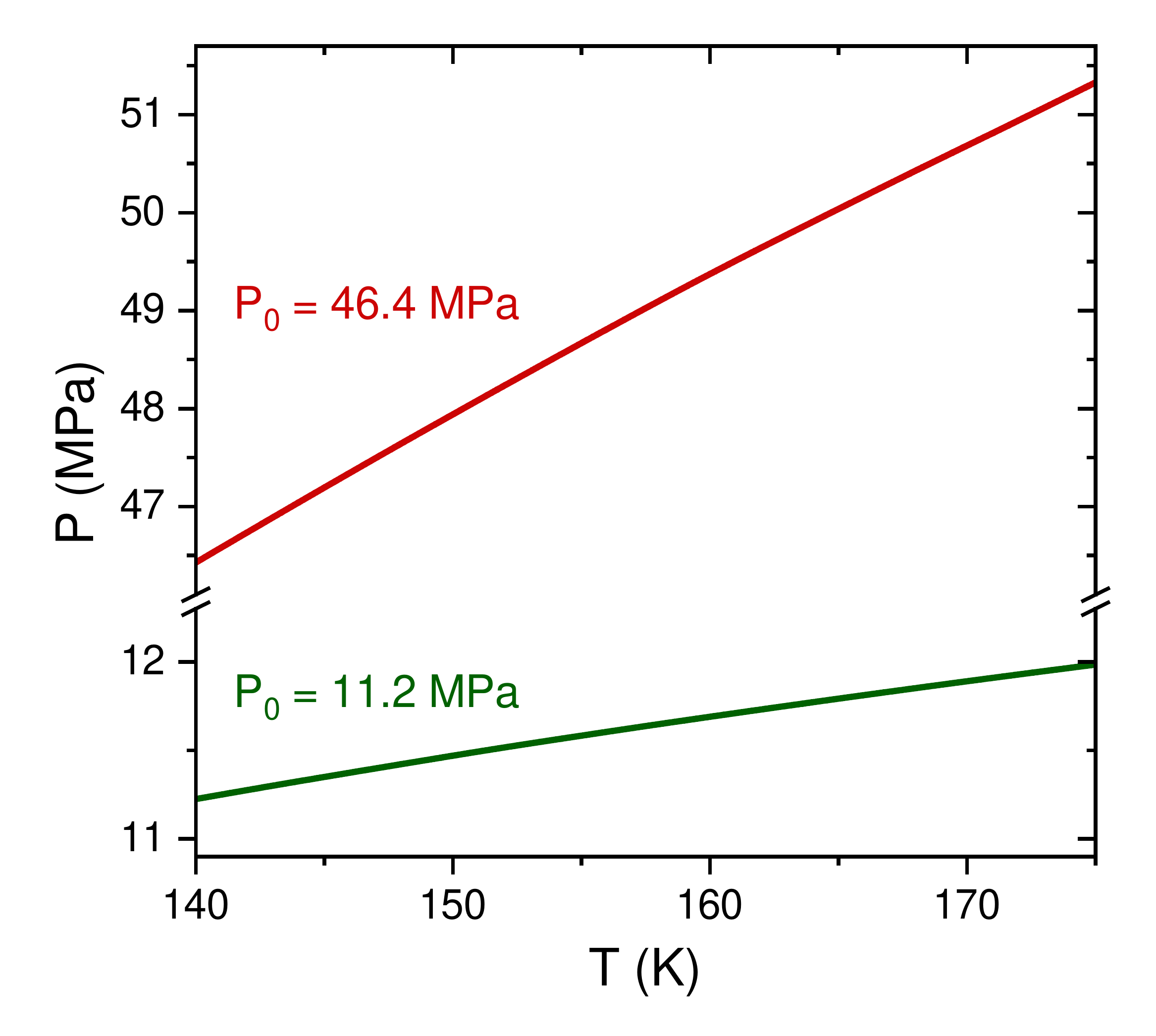}
	\caption{\label{fig:test_operation_A}Experimental data showing the variation of pressure during $T$-sweeps for starting values $P_0 = 11.2$\,MPa (green) and $P_0 = 46.4$\,MPa (red). In these experiments both setups were interconnected and connected to the compressor unit as described in chapter \ref{sec:OperationA} and Fig.\,\ref{fig:operation_A}. The $T$-sweeps were performed in both setups simultaneously, using heating rates $q_1 = q_2 = 1.5$\,K/h, while measuring the resulting variations in $P(T)$.  }
\end{figure}

This process in its idealized form is shown in Fig.\,\ref{fig:operation_A}\,\cyan{(a)}. In figure\,\ref{fig:test_operation_A} we show the variation of pressure, $P(T)$, during temperature sweeps with the initial pressure set to $P_0$ as indicated in the figure. The conditions apply to typical thermal expansion experiments aiming at measuring relative length changes as a function of temperature over an extended range of temperatures, while attempting to maintain $P \approx$  const. conditions. The warming rate chosen of $q_1 = q_2 = 1.5$\,K/h for the sample- and reference system 
ensures thermal equilibrium of all components inside the IVCs. Both pressure cells were interconnected by opening valves (IV) and (V), and connected to the compressor unit via an open valve (III). The figure demonstrates that for the experiment at lower pressure around 11\,MPa (green) the increase in temperature from $T_1 \sim 140$\,K to $T_2 \sim 175$\,K results in a moderate change of pressure with $\vert \Delta P \vert /P\sim5$\%. However, on increasing the target pressure to around 46\,MPa (red), the pressure change is significantly higher reaching values of $\vert \Delta P \vert /P\sim 10$\% upon traversing the same temperature window. As mentioned in section \ref{sec:OperationB}, this is due to the fact that at pressures $P>30$\,MPa the compressor stages have to be activated, which leads to a reduction in the remaining volume of the pressurized system. These two examples indicate that the performance in terms of pressure stability depends strongly on the chosen process parameters. In this mode of operation, corresponding to section \ref{sec:OperationA}, $P\approx$ const. conditions can only be realized at a sufficiently low pressure (due to the use of a larger gas volume) and/or for traversing a sufficiently narrow temperature interval.


\subsection{Pressure sweeps at $T \approx$ const. conditions } \label{sec:Psweep_at_T_const}
\begin{figure} [b]
	\centering
	\includegraphics[width=0.48\textwidth]{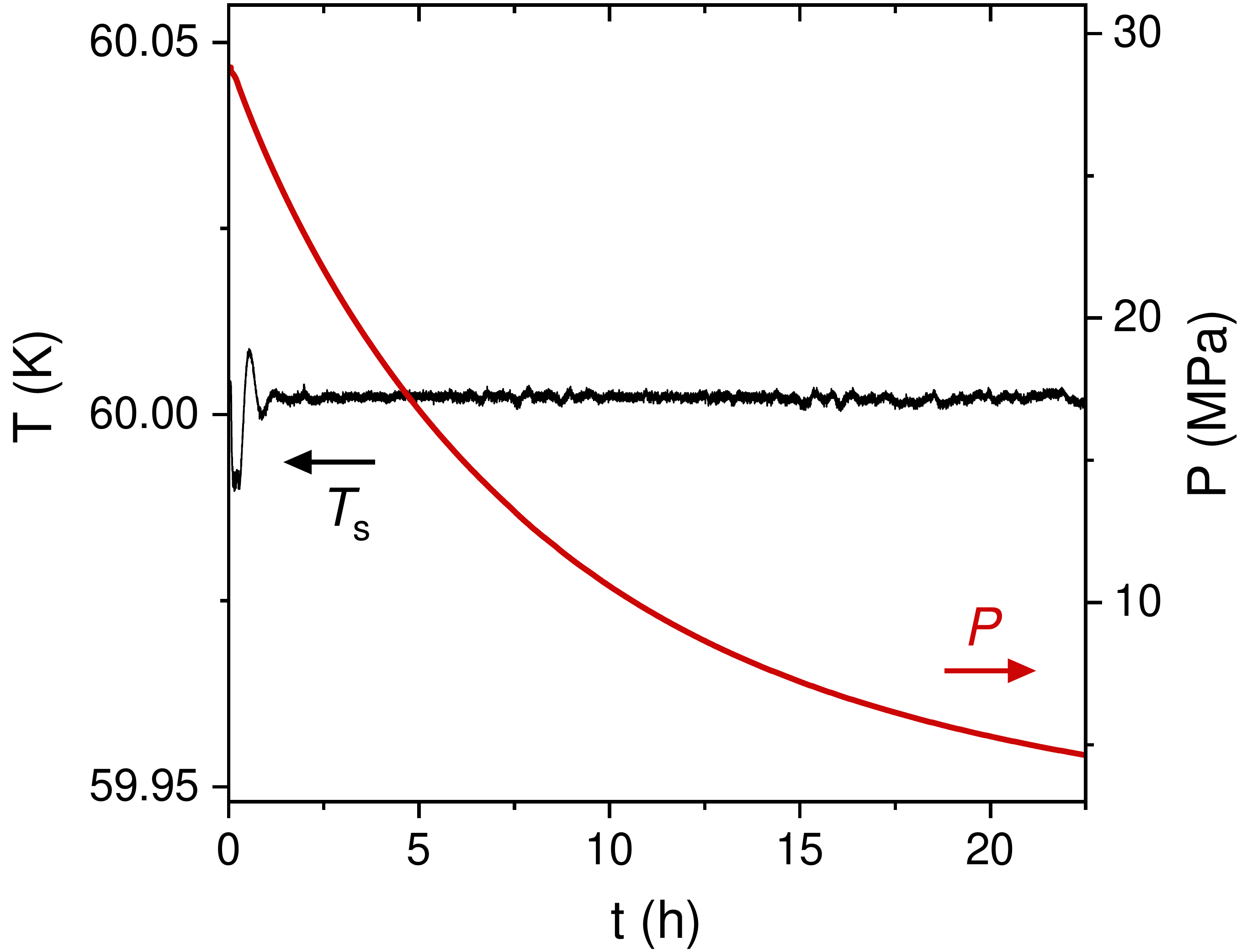}
	\caption{\label{fig:test_operation_B}Experimental data taken in the sample setup for temperature $T_\mathrm{s}$ (black, left scale) and pressure $P$ (red, right scale) as a function of time during a pressure sweep with intended $T\approx$ const. conditions. In this experiment both setups were interconnected and connected to the compressor unit as described in Fig.\,\ref{fig:operation_A}\,\cyan{(b)}. The pressure sweeps were performed simultaneously in both setups by slightly opening valve (VI) and releasing the pressure into the recovery line while measuring the associated temperature changes $T(t)$.}
\end{figure}
In order to perform pressure sweeps at $T \approx$ const. conditions, corresponding to a realization of the process shown in Fig.\,\ref{fig:operation_A}\,\cyan{(b)}, special measures would be required for ensuring a smooth, approximately $t$-linear change of pressure. In principle, this could be realized to some extent by using a controllable metering valve (VI) which would allow for a precise control of the flow rate by which the gas is released from the pressure cells. In the present system, however, which lacks such a metering valve, the flow rate can be adjusted by slightly opening valve (VI) manually. An example of the resulting pressure change as a function of time is depicted in Fig.\,\ref{fig:test_operation_B}. It shows a relatively rapid initial drop in $P(t)$, which turns into an approximately exponential decrease on a longer time scale. Apart from the initial phase of this process ($t < 3$\,h) , where the change in pressure is rather large $\vert \Delta P(t) \vert /\Delta t \geq 2$\,MPa/h and accompanied by some temperature instabilities, a smooth, approximately exponential variation in $P(t)$ can be realized in this way while keeping the variations in temperature small $\vert \Delta T \vert /T < 0.03$\,\%. Thus, this mode of operation is useful for experiments where decreasing pressure is desired/sufficient and the pressure range to be covered is not too large, typically $\vert \Delta P\vert < 25$\,MPa. As mentioned above, however, this mode of operation cannot be used for pressure sweeps with increasing pressure.


\subsection{Temperature sweeps with improved stabilization of pressure } \label{sec:Tsweep_at_P_const_improved}

\begin{figure} [h!]
	\centering
	\includegraphics[width=0.48\textwidth]{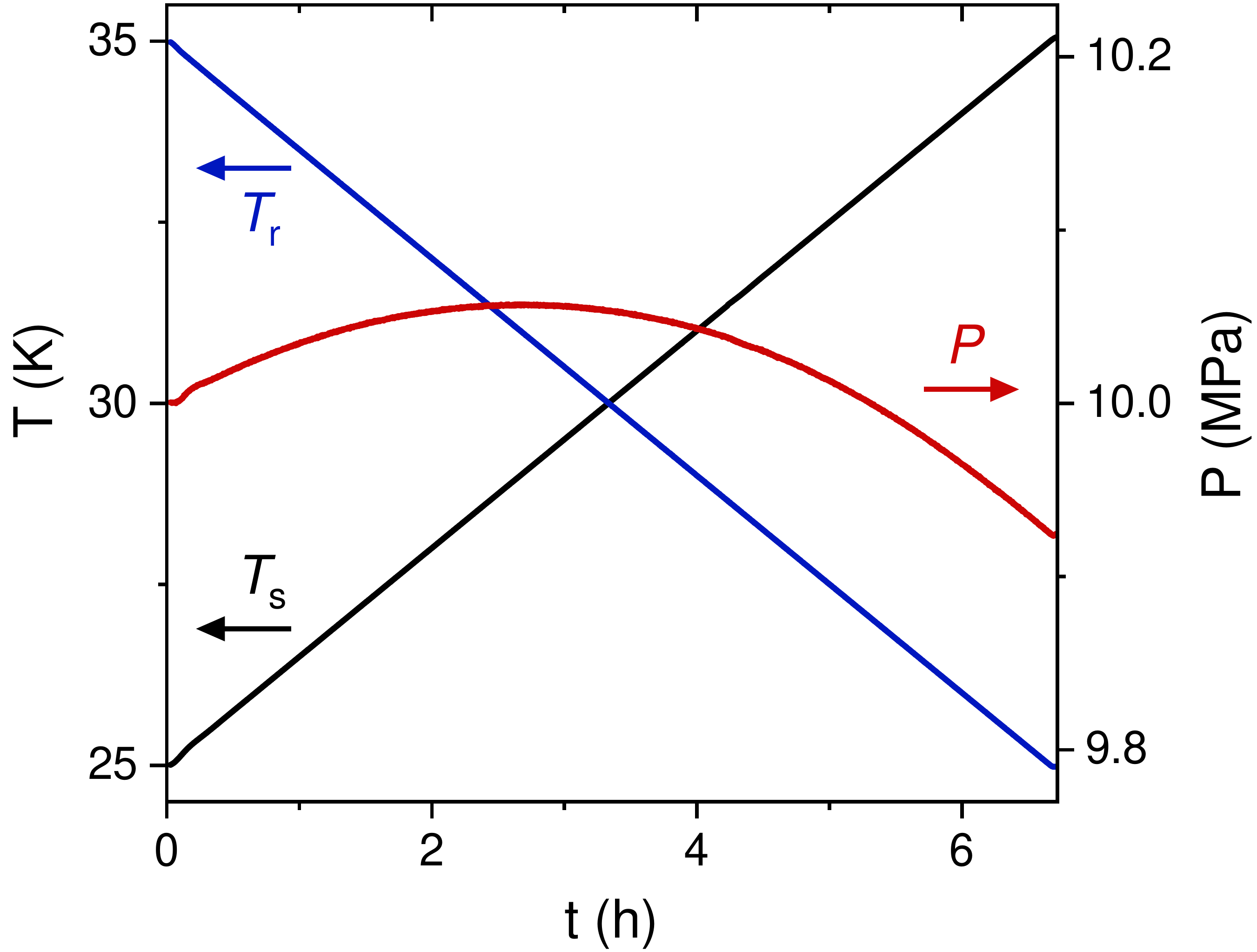}
	\caption{\label{fig:test_operation_C} Experimental data for the evolution of the sample temperature $T_\mathrm{s}$ (black)- and reference temperature $T_\mathrm{r}$ (blue, left scale) as well as the sample pressure (red, right scale) as a function of time during a temperature sweep with intended $P\approx$ const. conditions. In this experiment, both setups were interconnected and connected to the compressor unit as described in Fig.\,\,\ref{fig:operation_B}. The temperature sweeps were performed simultaneously in both setups using rates $q_1 = - q_2 = 1.5$\,K/h, while the associated pressure changes $P(t)$ were measured.}
\end{figure}

Figure \ref{fig:test_operation_C} shows the results of a temperature sweep experiment at $P\approx $ const., as sketched in Fig.\,\ref{fig:operation_B}, aiming at a high degree of pressure stabilization. In the experiment the sample cell was warmed from $T_1 = 25$\,K to $T_2 = 35$\,K with a heating rate $q_1 = 1.5$\,K/h while simultaneously cooling the reference cell with $q_2 =  - q_1$. The target pressure was $\sim 10$\,MPa. The observed variation of pressure, which qualitatively conforms to the simulations shown in Fig.\,\ref{fig:operation_B_model}\,\cyan{(a)}, reveal a high degree of stability of $\vert \Delta P\vert /P < 0.7$\,\%. For $P(t)$ we observe a dome-like shape, however, with some asymmetry. 
We assign this fact to small differences in the volumes of the pressure components for the sample (V$_{\mathrm{s}}$)- and reference (V$_{\mathrm{r}}$) setups. Indeed, accurate measurements of the involved volumes reveal $V_{\mathrm{s}}/V_{\mathrm{r}}\approx0.93$, which can account for this asymmetry.


\subsection{Pressure sweeps at an adjustable sweep rate } \label{sec:Psweep_at_adjustable_rate}
\begin{figure} [!b]
	\centering
	\includegraphics[width=0.48\textwidth]{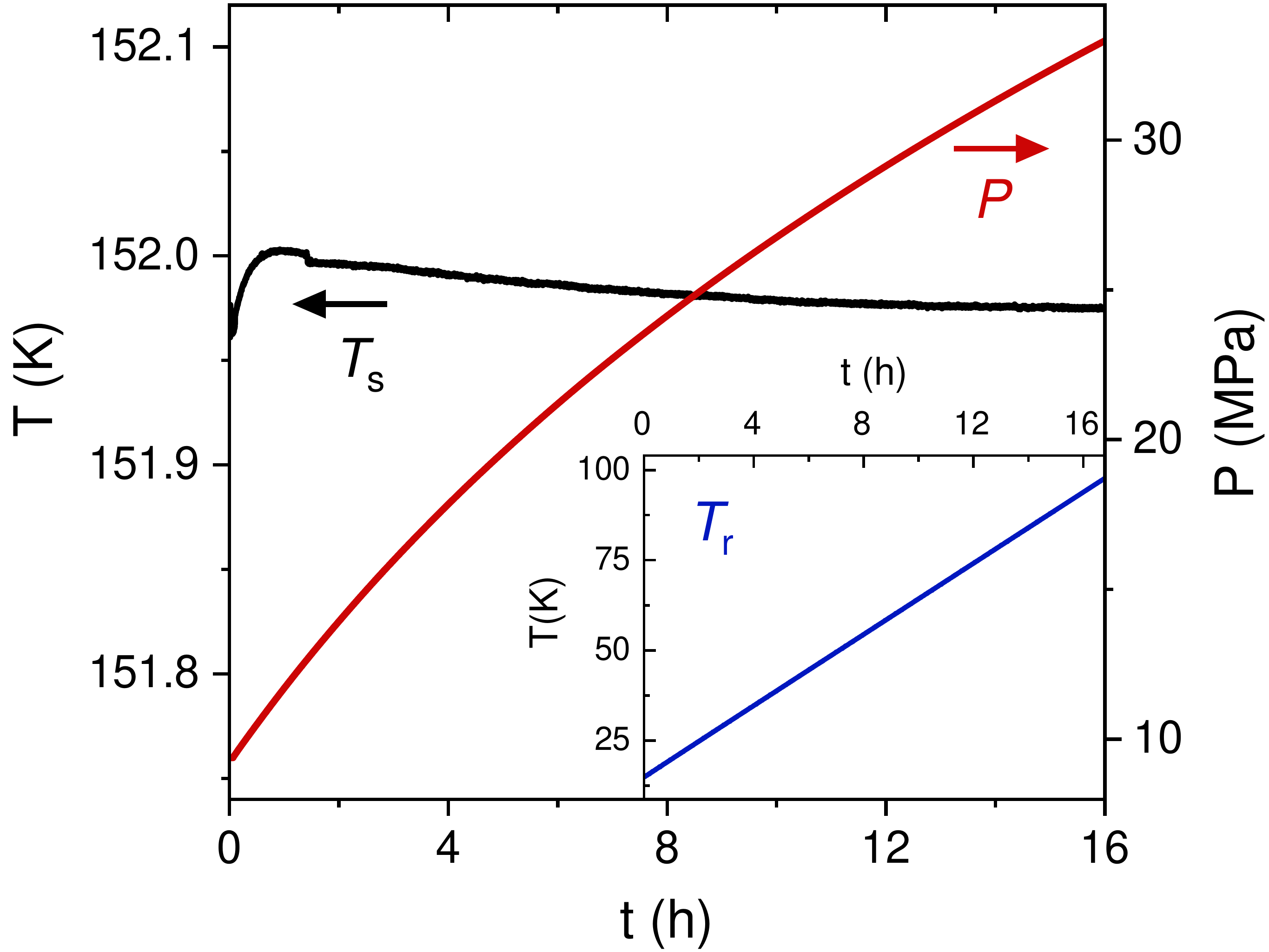}
	\caption{\label{fig:test_operation_D}Experimental data for the sample temperature $T_\mathrm{s}$ (black, left scale) and pressure $P$ (red, right scale) as a function of time for a pressure sweep with intended $T\approx$ const. conditions. In this experiment both setups were interconnected and connected to the compressor unit as described in Fig. 6 (a). The  pressure sweep was performed through heating the reference setup at a rate $q_2 = 5$\,K/h (inset), while measuring the accompanied variations in $P(t)$ in the sample pressure cell. Prior to the sweep the system was loaded with a starting pressure of $\sim10$\,MPa.}
\end{figure}
In Fig. \ref{fig:test_operation_D} we show results for a pressure sweep experiment performed as sketched in Fig.\,\ref{fig:operation_C}\,\cyan{(a)}. In this experiment, it was intended that the temperature of the sample setup,  $T_{\mathrm{s}}$, remains constant while the temperature of the reference system, $T_{\mathrm{r}}$, was varied by a rate of $q_2=5$\,K/h to induce a smooth increase of pressure in the whole system. Prior to the pressure sweep the system was loaded with a starting pressure of $P_1\sim10$\,MPa. By raising $T_{\mathrm{r}}$ from 15\,K to 100\,K the pressure increases smoothly from  $P_1\sim10$\,MPa to  $P_2\sim33$\,MPa. While in the initial phase of the sweep the pressure increase occurred at a rate $\Delta P/\Delta t \approx2.4$\,MPa/h, the rate decreased with time and reached a value of $\Delta P/\Delta t \approx1$\,MPa/h towards the end of the sweep. Apart from a small instability of $T_{\mathrm{s}}$ in the initial phase of the process ($t < 1$\,h), $T_{\mathrm{s}}$ remained constant within $\vert \Delta T\vert < 0.025$\,K, corresponding to $\vert \Delta T\vert/T < 0.02$\,\%. 
\begin{figure*} [!t]
	\centering
	\includegraphics[width=0.86\textwidth]{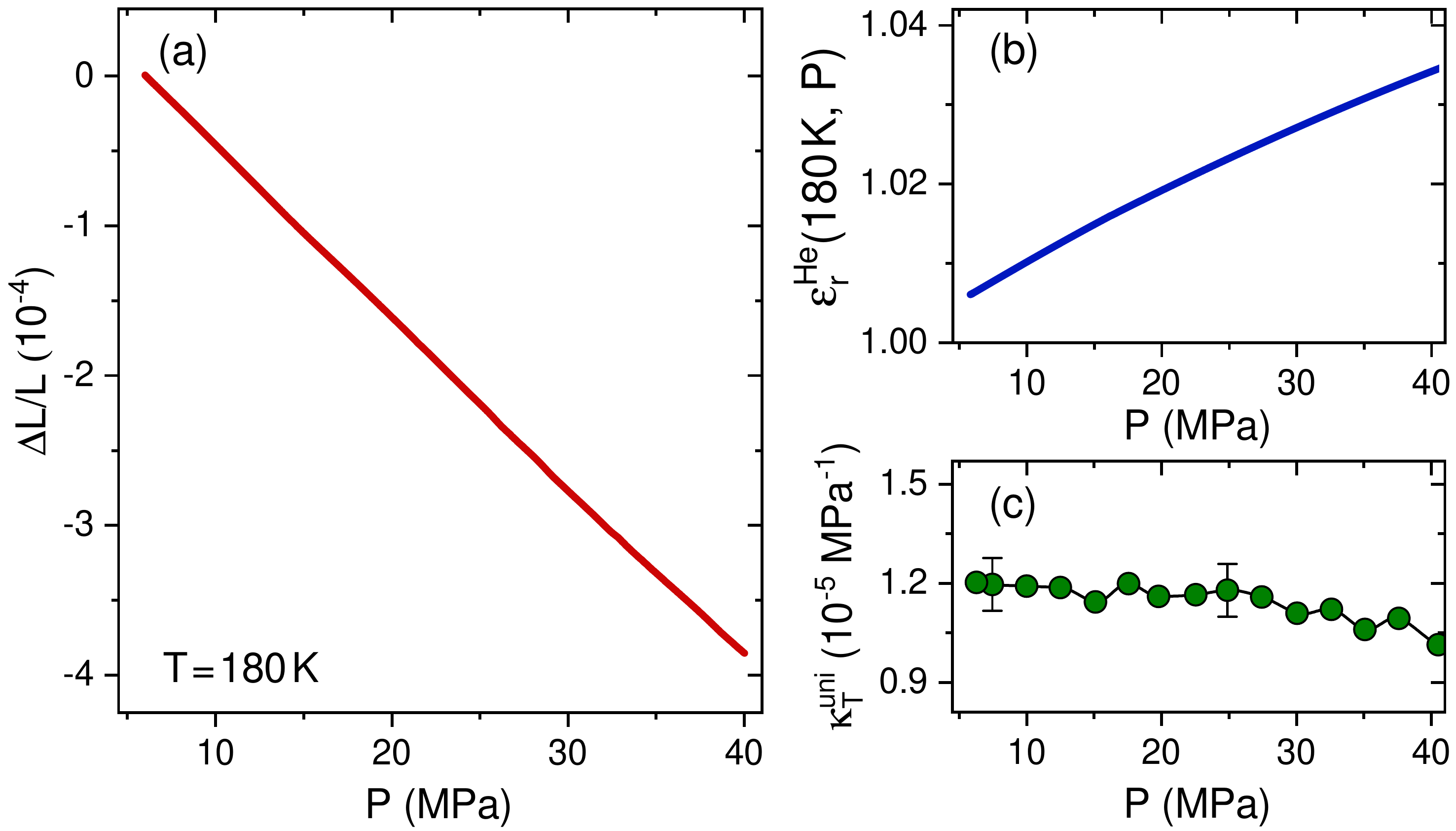}
	\caption{\label{fig:results_NaCl}(a) Relative length change $(\Delta L(P)/L)_T$ as a function of pressure $P$ at $T = 180$\,K for single crystalline NaCl with $L(300\,\mathrm{K}) = 6.8$\,mm. (b) Results of the dielectric constant of helium $\varepsilon_{\mathrm{r}}^{\mathrm{He}}(180\,\mathrm{K},P)$ derived from simultaneous measurements along the same $T$-$P$ trajectory in the reference setup. (c) Uniaxial isothermal compressibility, $\kappa_T^{\mathrm{uni}}$, of NaCl, derived as described in the text. Representative error bars resulting from the statistical treatment of the raw data and the estimated systematic uncertainties.}
\end{figure*}

The induced pressure changes depend sensitively on the parameters chosen in the reference system. In order to generate a high increase in pressure, a low starting temperature should be selected. Since a $t$-linear heating in the reference cell does not result in a $t$-linear increase of pressure in the sample system, a suitable, non-linear $T_{\mathrm{r}}$–$t$ profile can be chosen to generate an approximately $t$-linear increase in $P(t)$ if required.


\section{MEASUREMENTS OF RELATIVE LENGTH CHANGES (\textbf{$\Delta$}L/L)$_{\textbf{\textit{T}}}$ AS A FUNCTION OF PRESSURE ON SODIUM CHLORIDE} \label{sec:NaCl}

In order to test the performance of the setup, measurements of relative length changes $(\Delta L/L)_{T\sim\mathrm{const}}$  were conducted as a function of pressure on a single crystal of NaCl. This material was selected for these test studies due to its relatively high compressibility and its simple (cubic) structure. These facts along with the lack of any isomorphic transition up to 30\,GPa \cite{Johnson1966, Bassett1968} make this material a frequently used pressure-calibration standard \cite{Decker1965, Decker1971, Brown1999}. Due to the large amount of experimental data on this material, there have been various approaches to derive the equation of states, see Refs.\,\onlinecite{Decker1965, Decker1971, Brown1999, Dorogokupets2002, Sumita2013}. 

While the majority of studies were focusing on high pressures $P \geq 100$\,MPa and temperatures $T\geq300$\,K \cite{Bridgman1945, Barsch1967, Spetzler1972, Yamamoto1987, Chhabildas1976,  Boehler1980, Wang1996}, much less is known for the $P$-$T$ range of interest here, i.e., for pressures $P<100$\,MPa and temperatures $T<300$\,K. Nevertheless, the available data allow for an extrapolation into the low-$P$ range with sufficient accuracy, yielding an isothermal volume compressibility at $T=300$\,K of $\kappa _T = (4.20 \pm 0.01) \cdot 10^{-5}$\,MPa$^{-1}$, see Refs.\,\onlinecite{Chhabildas1976, Decker1972} and references cited therein. In addition, an extrapolation of the isothermal bulk modulus for lower temperatures can be obtained from Refs.\,\onlinecite{Yamamoto1987, Wang1996, Vinet1987}, which is useful for our purposes. 

In Fig.\,\ref{fig:results_NaCl}\,\cyan{(a)} we show results of relative length changes $(\Delta L/L)_T$ as a function of pressure for 6\,MPa $\leq P \leq 40$\,MPa at a constant temperature $T = 180$\,K taken on a single crystalline sample of NaCl. The sample length at 300\,K was $L = 6.8$\,mm. The experiment was performed in a mode of operation as described above in paragraph \ref{sec:Psweep_at_T_const}, i.e., both setups were interconnected and connected to the compressor unit as described in Fig.\,\ref{fig:operation_A}\,\cyan{(b)}. $P$-sweeps were performed simultaneously in both setups by slightly opening valve (VI) and releasing the pressure into the recovery line. This resulted in an initial sweep rate of $\vert \Delta P \vert /\Delta t\sim 2$\,MPa/hour for $P\leq 40$\,MPa, cf. Fig.\,\ref{fig:test_operation_C}. During the $P$-sweep the variation in temperature was $\vert \Delta T \vert \leq 1$\,K. The experiment allowed for measurements of $(\Delta L/L)_{T\approx 180\, \mathrm{K}}$ in the sample setup while simultaneously measuring the dielectric constant of helium \eps\ for the same $T$-$P$ trajectory. The obtained results for the dielectric constant of helium \eps\ are shown in Fig.\,\ref{fig:results_NaCl}\,\cyan{(b)} for the corresponding pressure range. These \eps\ data were used to transform the measured capacitance changes $(\Delta C/C)_{T \approx 180\,\mathrm{K}}$ into relative length changes $(\Delta L/L)_{T\approx 180\, \mathrm{K}}$ by using Eq.\,\ref{eq:capacitance_pressure}, see also Ref.\,\onlinecite{Manna2012}. The data in Fig.\,\ref{fig:results_NaCl}\,\cyan{(a)} yield a smooth decrease in $(\Delta L(P)/L)_T$ the slope of which gradually decreases with increasing pressure, indicating a hardening of the material. The high relative resolution of the measurement becomes even more clear by looking at the uniaxial compressibility $\kappa_T^{\mathrm{uni}}= -1/L\cdot (\partial L/ \partial P)_T$, approximated by $\kappa_T^{\mathrm{uni}} \approx -1/L \cdot (\Delta L/\Delta P)_T$, shown in Fig.\,\ref{fig:results_NaCl}\,\cyan{(c)}. For determining the differential quotient, the $\Delta L/L$-data were divided into equal pressure intervals of width 2.5\,MPa in each of which a mean slope is determined by linear regression. The data, in the limit $P \rightarrow 0$, correspond to a volume compressibility of $\kappa_T^{\mathrm{vol}}= 3 \cdot \kappa_T^{\mathrm{uni}}=(3.60\pm 0.24)\cdot 10^{-5}$\,MPa$^{-1}$. This value can be compared with the extrapolated value (for $T=180$\,K) of the isothermal bulk modulus $B_T=1/\kappa_T^{\mathrm{vol}}$ of 25.6\,GPa reported in Refs.\,\onlinecite{Yamamoto1987, Wang1996}, corresponding to $\kappa_T^{\mathrm{vol}} = 3.9\cdot 10^{-5}$\,MPa$^{-1}$. Note that the calculation of the equation of state, which forms the basis for the latter value, is determined from experimental quantities associated with relative errors of typically a few percent, cf.\,Ref.\,\onlinecite{Yamamoto1987}. Even though the present apparatus has been designed aiming at a high relative resolution, the above comparison of compressibility data for NaCl demonstrates a satisfactory absolute accuracy.


\section{Conclusion} \label{sec:Conclusion}

We have realized an advanced technique for measuring relative length changes of mm-sized samples under control of temperature and helium-gas pressure. The key elements of the apparatus include a room-temperature helium-gas reservoir, which is connected to two He-bath cryostats each of which is equipped with a pressure cell and a capacitive dilatometer. The system enables to run different, well-controlled $T(t)$ and $P(t)$ profiles which can be identical for both setups or individually customized for generating a desired $T$-$P$ trajectory in the sample setup. As compared with an earlier version of this apparatus \cite{Manna2012}, consisting of a single cryostat housing the sample setup, this extended version offers new possibilities for performing high-resolution measurements of relative length changes under well-controlled conditions. The system is distinguished by the following features: 1) It enables to significantly increase the maximum accessible pressure, $P_{\mathrm{max}}$, being no longer limited by the external pressure reservoir. 2) The unwanted change of pressure during a temperature sweep has been drastically reduced. 3) Pressure sweeps can be performed both with decreasing and increasing pressure for a wide range of sweeping rates. 4) The setup allows for a simultaneous measurement of \eps\ along the same $T$-$P$ trajectory as used for taking the $\Delta L(T,P)/L$ data. This setup is well suited for detailed investigations of lattice effects and their coupling to other degrees of freedom in certain ranges of the $T$-$P$ phase diagram.


\begin{acknowledgments}
We acknowledge support by the Deutsche Forschungsgemeinschaft (DFG, German Research Foundation) through TRR 288 - 422213477 (projects A01 and A06) and K.D. Luther for providing us with the NaCl crystal.
\end{acknowledgments}

\section*{AUTHOR DECLARATIONS}

\subsection*{Conflict of Interest}

The authors have no conflicts to disclose.

\section*{Data Availability Statement}

The data that support the findings of this study are available from the corresponding author upon reasonable request.



\nocite{*}
\bibliography{RSI_paper_literature}

\end{document}